\documentclass[sigconf]{acmart}

\hypersetup{bookmarks=false}
\PassOptionsToPackage{bookmarks=false}{hyperref}
\usepackage[table]{xcolor}
\usepackage[svgnames]{xcolor}
\usepackage{multirow}
\usepackage{booktabs} 
\usepackage{amsmath}
\usepackage{graphicx}
\usepackage[belowskip=-12pt]{caption}
\usepackage{rotating}
\usepackage{tabularx}
\usepackage{subfig}
\usepackage{wrapfig}
\usepackage{listings}
\usepackage{color, colortbl}
\usepackage{balance} 
\usepackage{lscape}
\usepackage{xspace}
\usepackage{framed}
\usepackage{syntax}
\usepackage{soul}
\usepackage{flushend}
\usepackage[tikz]{bclogo}
\usepackage{makecell}
\definecolor{codegreen}{rgb}{0,0.6,0}
\definecolor{codegray}{rgb}{0.5,0.5,0.5}
\definecolor{codepurple}{rgb}{0.58,0,0.82}
\definecolor{backcolour}{rgb}{0.95,0.95,0.92}
\definecolor{Gray}{gray}{0.1}
\lstdefinestyle{mystyle}{
	backgroundcolor=\color{backcolour},   
	commentstyle=\color{codegreen},
	keywordstyle=\color{magenta},
	numberstyle=\tiny\color{codegray},
	stringstyle=\color{codepurple},
	basicstyle=\scriptsize,
	breakatwhitespace=false,         
	breaklines=true,                 
	captionpos=b,                    
	keepspaces=true,                 
	numbers=left,                    
	numbersep=5pt,                  
	showspaces=false,                
	showstringspaces=false,
	showtabs=false,                  
	tabsize=2
}

\definecolor{diffstart}{named}{Grey}
\definecolor{diffincl}{named}{ForestGreen}
\definecolor{diffrem}{named}{OrangeRed}

\lstdefinelanguage{Pythonna}{%
	language     = python,
	morekeywords = {to_categorical, flow_from_directory,pad_sequences,load_image},
	morecomment=[f][\color{diffstart}]{@@},
	morecomment=[f][\color{diffincl}]{+\ },
	morecomment=[f][\color{diffrem}]{-\ }	
}

\lstdefinestyle{customc}{
	belowcaptionskip=1\baselineskip,
	breaklines=false,
	frame= single,
	breaklines = true,
	xleftmargin=\parindent,
	language= Pythonna,
	showstringspaces=false,
	basicstyle=\footnotesize\ttfamily,
	keywordstyle=\bfseries\color{green!40!black},
	commentstyle=\itshape\color{purple!40!black},
	identifierstyle=\color{blue},
	stringstyle=\color{codegreen},
	backgroundcolor=\color{gray!4}
}

\newcommand{\centerhfill}[1][\quad]{\hspace{\stretch{0.5}}#1\hspace{\stretch{0.5}}}
\graphicspath{{./figures/}}
\newcommand{\figref}[1]{Fig.~\ref{#1}}
\newcommand{\tabref}[1]{Table~\ref{#1}}

\newcommand{\etal}{{\em et al.}\xspace}

\newcommand{\code}[1]{{\texttt{\footnotesize{#1}}\xspace}}

\newcommand{\sof}{\textit{Stack Overflow}\xspace}
\newcommand{\caffe}{\textit{Caffe}\xspace}

\newcommand{\keras}{\textit{Keras}\xspace}

\newcommand{\scikit}{\textit{scikit-learn}\xspace}
\newcommand{\tensor}{\textit{Tensorflow}\xspace}
\newcommand{\theano}{\textit{Theano}\xspace}
\newcommand{\torch}{\textit{Torch}\xspace}

\newcommand{\gh}{\textit{GitHub}\xspace}
\newcommand{\paddle}{\textit{paddle}\xspace}

\newcounter{rqs}
\stepcounter{rqs}

\newcommand{\sbug}{\textit{320}\xspace}
\newcommand{\gbug}{\textit{347}\xspace}

\newcounter{NumObservations}
\stepcounter{NumObservations}
\definecolor{shadecolor}{rgb}{.9,.9,.9}

\newcommand{\finding}[1]{
	\begin{bclogo}[couleur= blue!5, epBord= 1, arrondi=0.1, logo=\bclampe,marge= 2, ombre=true, blur, couleurBord=blue!60, tailleOndu=3, sousTitre ={\em #1}]{Finding \arabic{NumObservations} $\Rightarrow$ } 
		
	\end{bclogo}
	\stepcounter{NumObservations}
}

\AtBeginDocument{%
  \providecommand\BibTeX{{%
    \normalfont B\kern-0.5em{\scshape i\kern-0.25em b}\kern-0.8em\TeX}}}

\copyrightyear{2020}
\acmYear{2020}
\setcopyright{acmcopyright}
\acmConference[ICSE '20]{42nd International Conference on Software Engineering}{May 23--29, 2020}{Seoul, Republic of Korea}
\acmBooktitle{42nd International Conference on Software Engineering (ICSE '20), May 23--29, 2020, Seoul, Republic of Korea}
\acmPrice{15.00}
\acmDOI{10.1145/3377811.3380378}
\acmISBN{978-1-4503-7121-6/20/05}

\usepackage{bookmark}
\makeatletter
\renewcommand{\toclevel@section}{100}
\renewcommand{\toclevel@subsection}{101}
\renewcommand{\toclevel@subsubsection}{102}
\renewcommand{\toclevel@paragraph}{103}
\renewcommand{\toclevel@subparagraph}{104}
\begin{document}

\title{Repairing Deep Neural Networks: Fix Patterns and Challenges}
\author{Md Johirul Islam}
\email{mislam@iastate.edu}
\affiliation{%
	\institution{Dept. of Computer Science, Iowa State University}
	\streetaddress{226 Atanasoff Hall}
	\city{226 Atanasoff Hall, Ames}
	\state{IA}
	\postcode{50011}
	\country{USA}
}

\author{Rangeet Pan}
\email{rangeet@iastate.edu}
\affiliation{%
	\institution{Dept. of Computer Science, Iowa State University}
	\streetaddress{226 Atanasoff Hall}
	\city{226 Atanasoff Hall, Ames}
	\state{IA}
	\postcode{50011}
	\country{USA}
}

\author{Giang Nguyen}
\email{gnguyen@iastate.edu}
\affiliation{%
	\institution{Dept. of Computer Science, Iowa State University}
	\streetaddress{226 Atanasoff Hall}
	\city{226 Atanasoff Hall, Ames}
	\state{IA}
	\postcode{50011}
	\country{USA}
}

\author{Hridesh Rajan}
\email{hridesh@iastate.edu}
\affiliation{%
	\institution{Dept. of Computer Science, Iowa State University}
	\streetaddress{226 Atanasoff Hall}
	\city{226 Atanasoff Hall, Ames}
	\state{IA}
	\postcode{50010}
	\country{USA}
}


\begin{abstract}
Significant interest in applying Deep Neural Network (DNN) 
has fueled the need to support engineering of software that 
uses DNNs.
Repairing software that uses DNNs is one such unmistakable SE
need where automated tools could be beneficial; however, 
we do not fully understand challenges to repairing and 
patterns that are utilized when manually repairing DNNs.
What challenges should automated repair tools address? 
What are the repair patterns whose automation could help 
developers? Which repair patterns should be assigned a 
higher priority for building automated bug repair tools?
This work presents a comprehensive study of bug fix patterns
to address these questions. 
We have studied 415 repairs from \sof and 555 repairs from
\gh for five popular deep learning libraries \caffe, \keras, 
\tensor, \theano, and \torch to understand challenges in 
repairs and bug repair patterns.
Our key findings reveal that 
DNN bug fix patterns are distinctive compared to traditional bug fix patterns;  
the most common bug fix patterns are fixing data dimension and neural network connectivity; 
DNN bug fixes have the potential to introduce adversarial vulnerabilities;
DNN bug fixes frequently introduce new bugs; and 
DNN bug localization, reuse of trained model, and coping with frequent releases 
are major challenges faced by developers when fixing bugs. 
We also contribute a benchmark of 667 DNN (bug, repair) instances.
\end{abstract}

%


\maketitle

\section{Introduction}
\label{sec:intro}
The availability of big data has fueled the emergence
of deep neural networks (DNN).
A DNN consists of a set of layers. Each layer contains a set
of nodes collecting inputs from the previous layer and feeding the output 
to nodes in the next layer via a set of weighted edges.
These weights are adjusted using examples, called training data, and set 
to values that minimize the difference between actual outputs of the DNN and 
expected outputs measured using an objective function called loss function.
The availability of big data has made it possible to accurately adjust  
weights for DNNs containing many layers.
Thus, many software systems are routinely utilizing DNNs.
SE for DNNs has thus become important.

A significant SE problem in the software that uses DNNs is the presence of bugs.
What are the common bugs in such software? How do they differ? 
Answering these questions has the potential to fuel SE research 
on bug detection and repair for DNNs.
Fortunately, recent work has shed some light on this issue. 
Zhang~\etal~\cite{zhang2018empirical} have identified bug types, root
causes, and their effects in \tensor library for DNN.
Islam~\etal~\cite{islam2019comprehensive} have studied an even larger set of
libraries including \caffe, \keras, \tensor, \theano, and \torch
to identify bug characteristics.
While prior work presents an initial study on repair patterns
for \tensor, these works have not focused on the characteristics of repairs.
Since repairing software that uses DNNs is an unmistakable SE
need where automated tools could be very helpful, fully 
understanding the challenges to repairing and patterns that 
are utilized when manually repairing bugs in DNNs is critical.
What challenges should automated repair tools address? 
What are the repair patterns whose automation could help developers? 
Which repair patterns should be prioritized?

Motivated by these questions, we conduct a comprehensive study of 
bug repair patterns for five DNN libraries 
\caffe, \keras, \tensor, \theano, and \torch.
We leverage the dataset of DNN bugs published by
Islam~\etal~\cite{islam2019comprehensive}
that consists of 415 bugs from \sof and 555 bugs from \gh.
We then collect the code snippets used to fix these bugs from
both \sof and \gh.
We then manually study these repairs and label them according to a 
classification scheme developed using the open coding approach. 
To study the fix in \sof we study the 
accepted answers and answers with score >= 5 
from \sof post that fixes the bug in the original post.
To study the bug fix patterns in \gh, we take the bug-fix 
commits in the dataset and study the code that is changed to fix the bug. 
If we do not find any fixes that match
our selection criteria
in \sof and relevant fix in \gh 
we discard those bugs.
In total, we have studied \sbug bug fix codes in 
\sof and \gbug bug fix codes in \gh. 
We have also analyzed these bug fixes to answer the 
following research questions:
\begin{itemize}
	\item[RQ1] \textbf{(Common bug fix patterns)} What are the most common bug fix
	patterns?
	
	\item[RQ2] \textbf{(Fix pattern across bug types)} Are the bug fix patterns
	different for different bug types?
	
	\item[RQ3] \textbf{(Fix pattern across libraries)} Are the bug fix pattern
	different for different libraries?
	
	\item[RQ4] \textbf{(Risk in fix)} Does fixing a DNN bug introduces a new bug?
	
	\item[RQ5] \textbf{(Challenges)} What are the challenges in fixing DNN
	bugs?
\end{itemize}
Our key findings are as follows:
DNN bug fix patterns are distinctive compared to traditional bug fix patterns;  
the most common bug fix patterns are fixing data dimension and network connectivity; 
DNN bug fixes have the potential to introduce adversarial vulnerabilities~\cite{goodfellow2014explaining}; 
DNN bug fixes frequently introduce new bugs; and 
DNN bug localization, reuse of trained model, and coping with frequent releases 
are major challenges faced by developers when fixing bugs. 
We also contribute a benchmark of 667
DNN (bug, repair) instances. 
This benchmark is also publicly accessible \cite{dataset}.

\section{Methodology}
\label{sec:methodology}

\subsection{Dataset}
In our study, we build on the bug dataset prepared by 
Islam~\etal~\cite{islam2019comprehensive} to collect and 
to prepare the dataset of bug fixes. 
The bug dataset contains $415$ bugs from \sof and $555$ 
bugs from \gh for 5 different deep learning libraries 
as shown in \tabref{tbl:dataset}. 

\begin{table}[h!t]
\centering
\caption{Summary of the bug repair dataset.}
\label{tbl:dataset}
\setlength{\tabcolsep}{3pt}
\rowcolors{2}{gray!25}{white}
	\begin{tabular}{| l | r | r | r | r |}
		\hline
	\rowcolor{gray!50}
   Library & \multicolumn{2}{c|}{\sof} & \multicolumn{2}{c|}{\gh} \\ \cline{2-5} 
   & Bugs \cite{islam2019comprehensive} & Fixes  (current)   & Bugs \cite{islam2019comprehensive}               & Fixes (current)  \\ \hline
    \caffe & 35                 & {\bf 27}                 & 26                 & {\bf 17} \\
    \hline
    \keras & 162                & {\bf 143}                & 348                & {\bf 167}  \\
    \hline
    \tensor & 166                & {\bf 118}                & 100                & {\bf 90} \\
    \hline
    \theano & 27                 & {\bf 15}                 & 35                 & {\bf 32} \\
    \hline
    \torch & 25                 & {\bf 17}                 & 46                 & {\bf 41} \\
    \hline
    \hline
    Total & 415                & {\bf 320}                & 555                & {\bf 347} \\
    \hline
	\end{tabular}
\end{table}

\textbf{Collecting \sof bug fixes: } 
To collect the bug fixes in \sof bug dataset, 
we study all the answers corresponding to the post ids 
in \sof bug dataset. 
If a post has accepted an answer with code, then we consider that code snippet as a
fix. 
If the accepted answer doesn't have code but describes
what needs to be fixed in the original bug we consider those as fix as well. 
If a bug post does not have an accepted answer but has an answer with 
>= 5 scores we consider them as fixes also as score 5 is considered as an acceptable quality metric in prior works \cite{islam2019comprehensive}.
Following this methodology, we were able to find 320 
fixes for 320 bug related posts in the \sof dataset. 

\textbf{Collecting \gh bug fixes: } 
To collect \gh bug fixes, we went to the link of the buggy code snippets in the
dataset. 
If the code snippet was fixed in a later revision, 
then we take those fixes. 
A single line may contain multiple bugs~\cite{islam2019comprehensive}.
A single bug fix commit might fix multiple bugs.
We consider them different fixes.
For example, in the same fix API name is updated from deprecated to a new
version and the dimension is also fixed. 
We consider them as two different fixes.
Some of the bugs are not yet fixed in the repositories and some repositories
have been made private or deleted
since the previous study. We omitted those bugs.
Following this methodology, we collected 347 bug fixes from \gh. 
%
\begin{table}[t]
\centering
\caption{Summary of the bug fix patterns.}
\label{tbl:fixpatterns}
\setlength{\tabcolsep}{3pt}
\rowcolors{2}{gray!25}{white}
	\begin{tabular}{|p{0.7in}|p{2.5in}|}
	\rowcolor{gray!50}
	\hline
   Bug \linebreak Fix Pattern & Definition \\  
    \hline
   Loss function & add, remove or replace the loss function. \\
    \hline
   Network \linebreak connection & change node connectivity in the DNN, e.g. change weights, remove edges, add backward propagation. \\
    \hline
   Add layer & add another layer to the DNN model\\
    \hline 
   Layer \linebreak dimension & change a layer's input and output size, e.g. to make it compatible with adjacent layers' dimension\\ 
    \hline 
    Data \linebreak dimension & align the input data's dimension with the layer dimension\\ 
    \hline 
    Accuracy \linebreak metric & replace the accuracy metric being used to measure the correctness of a model, often to match better\\
    \hline 
    Data type & change the type of data given as input to the DNN\\
    \hline 
    Activation & change the activation function used in the DNN\\
    \hline 
    Iterations & change the number of times the training would be done, e.g. modify batch size, epoch or add a loop\\
    \hline
    Versioning & adapt the code to the new version of the library\\
    \hline 
    API contract & fix API compositions so that the output of an API meets the preconditions of another API\\
    \hline 
    Data \linebreak wrangling & fix the form
     of the data for downstream operations without modifying its intent\\
    \hline 
    Monitor & add diagnostics code to monitor training\\
    \hline 
    Optimizer & change the optimization function used by the DNN\\
    \hline 
    Change neural \linebreak architecture & overhaul the design of the DNN's architecture 
                      including a new set of layers and hyperparameters, generally because changes above can't fix the bug\\
    \hline
	\end{tabular}
\end{table}

\subsection{Bug Fix Pattern Classification}
Next, we created a classification scheme to manually label the bug fix
dataset. 
We started with the classification scheme used
by Pan, Kim, and Whitehead~\cite{pan2009toward} and found that their classification scheme has
	28 non-ML bug fix categories and among them only 4 fix categories are applicable for the
	DNN-related fixes. Then, we used the
open coding approach to refine it to come with a pattern of 15 different
kinds of DNN-specific bug fix patterns. 
We conducted a pilot study where two Ph.D. students individually studied the
fixes to come up with a possible classification. 
Each student proposed a set of classes that were then reconciled during
an in-person meeting where all the authors were present. 
In the in-person meeting, the authors validated the classification schemes 
from two individual raters and updated the classification scheme based on 
the outcome of the reconciliation effort under the supervision of the moderator.
Our pilot study revealed that there are a number of unique bug fix patterns
in our DNN setting. 
Therefore, the classification from prior work had to be significantly modified.
The final classification is shown in \tabref{tbl:fixpatterns} and discussed
below.
\finding{We found that DNN bug fix patterns are 
	very different from traditional bug fix patterns such as~\cite{pan2009toward}. }
\subsubsection{Loss Function} 
This group of fixes is based on the addition, removal, or update of the loss
function during training. 
The loss function is a key parameter that helps the training process to identify
the deviation from the learned and actual examples. 
Different kind of problems demand a different loss function, 
e.g., cross-entropy loss is widely used in the classification problems whereas
mean square error loss (MSE) is mostly used for regression-based problems. 
Some problems ask for a custom loss function for better training result 
and we group this kind of fixes into this class.

\subsubsection{Network Connection}
This group of fixes changes the connection between nodes in the DNN.
A DNN is a graph, where edges are the weights and bias and nodes are the
elements of each layer. 
For example, in a dense layer, the weight edges are fully connected with 
the next layer and the dimension of the layer determines the number of 
nodes to be available in that layer.
Those bug fixes that reconfigure these connections for better results are 
classified in this category.  
The changes include change of weight, removing edges by pruning the network,
adding backward propagation, etc. 
\subsubsection{Add Layer}
In any classification based problem, there will be at least two layers in the
model, the input layer, and the output layer. To learn the features of the input, a
DNN frequently needs more intermediate layers (called hidden).
This group of fixes adds more layers to the DNN to improve performance.
Added layers can be dense, where two consecutive layers are fully 
connected, convolution layer, where convolution function has been 
applied to the input, dropout layer for reducing the overfitting, etc.
\subsubsection{Layer Dimension}
These fixes change the dimensions of the layers to make them compatible
with adjacent layers and input. 
\subsubsection{Data Dimension}
Data dimension related fix is similar to layer dimension, but it is related to
the input data rather than to the DNN layers. The dimension of the data needs
to be aligned with the DNN. This type of fix is mostly needed when the input
dimensions of the DNN and the data dimension do not match.
\subsubsection{Accuracy Metric}
To measure the correctness of a DNN, the accuracy metric is one of the key
parameters to be configured. 
The problem type has a huge influence on the type
of accuracy metric to be used, e.g., classification problems are judged using 
classification accuracy, F1 score or confusion matrix, but these metrics are 
unsuitable for assessing a regression-based model where logarithmic loss is more
suitable. 
\subsubsection{Data Type}
This group of fixes changes the data type of inputs to match the DNN's
expectation.
\subsubsection{Activation}
The activation function for a node in a layer of DNN maps
inputs to the output.
This group of fixes changes the activation function used in a layer to better
match the problem. 
\subsubsection{Iterations}
This group of fixes adjusts the number of times the training process will be run.

This is generally done to improve accuracy or to reduce overfitting. 
These fixes include changing batch size or epoch. 
In some cases, developers add a loop around the entire training process. 
%
\subsubsection{Versioning}
DNN libraries are being rapidly developed, and a number of releases are not 
backward compatible that breaks code. This group of fixes adapts a code to 
work with the new version of the DNN library. 
%
\subsubsection{API Contract}
When the output of a DNN API is fed to the input of another DNN API
operation, these two operations have to be compatible. 
This group of fixes adds adapters to fix incompatibilities between composed
operations. 
\subsubsection{Data Wrangling}
Data wrangling refers to changing the form of data without changing its intent.
It is generally done to fix the data for the downstream operations. 
This group of fixes adds data wrangling to fix a DNN, e.g. by data shifting, 
shuffle, etc. 
%
\subsubsection{Monitor}
The fixes in this category add code for diagnostics during the training process,
typically to print training statistics. 
This group of fixes do not repair
the flaw in the code, but they help to localize the bug.
\subsubsection{Optimizer}
This group of fixes modifies the optimization algorithms used by the DNN model.
The optimization algorithm, which is dependent on the problem,
determines the iterative process followed to improve the accuracy of the DNN
model. 
\subsubsection{Change Neural Architecture}
This group of fixes essentially re-do the DNN model because the initial model
was unsuitable. 
%

\subsection{Labeling}
For labeling, we used the classification scheme shown in \tabref{tbl:fixpatterns}. 
Two Ph.D. students with expertise in these DNN libraries were
requested to label the fixes according to the classification scheme. 
We held multiple training sessions to train the raters with the classification
scheme.
We used the Kappa coefficient \cite{viera2005understanding} to measure the agreement between the raters after
the labeling of every 100 bug fix patterns. 
We found that the Kappa coefficient was 82\% for the first 100 labelings, 
85\% for the second 100 labeling. 
This high value of the Cohen's Kappa coefficient indicates perfect agreement 
between the raters.
In the presence of a moderator, the repair patterns for which there was a label
conflict between the raters were reconciled. 
We adapted this methodology from \cite{islam2019comprehensive}. 
Following this strategy, we labeled all the fixes and reconciled the labeling
conflicts through moderated discussions. 
The Kappa score throughout the process was >85\% indicating a clear
understanding and perfect agreement. 
\section{Bug Fix Patterns}
\label{sec:pattern}

\begin{figure*}[!ht]
	\subfloat[Distribution of Bug Fix Patterns in \sof (Labels less than 3.1\% are hidden)\label{fig:bugpatternso}]{%
		\includegraphics[, trim={2cm 8cm 0cm 8cm},width=0.4\textwidth]{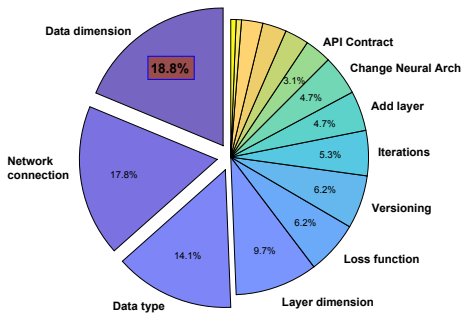}
	}
	\hfill
	\subfloat[Distribution of Bug Fix Patterns in \gh (Labels less than 3.2\% are hidden)	\label{fig:bugpatterngit}]{%
		\includegraphics[, trim={2cm 8cm 0cm 8cm},width=0.4\textwidth]{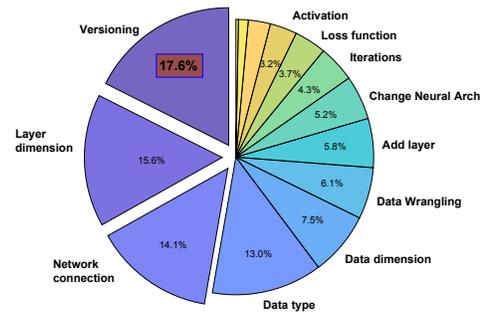}
	}
	\caption{Bug fix pattern distribution}
	\label{fig:bugpattern}
\end{figure*}

In this section, we explore the answer to RQ1 to understand what are the 
most common bug fix patterns in DNN. 
To answer RQ1, we take the labeled dataset and statistical distribution 
of the bug fix patterns across different categories. 
We also analyze the source code and diffs for those fixes to 
understand the challenges underlying those patterns. 
\figref{fig:bugpattern} shows the distribution of different bug fix patterns in \sof and \gh.


\begin{table}
	\vspace{16pt}
	\centering
	\caption{Bug Fixes in \sof (SO) and \gh (GH)}
	\label{tbl:gitsobugs}
	\setlength{\tabcolsep}{2pt}
	\footnotesize
	\renewcommand{\arraystretch}{0.5}
	\rowcolors{2}{gray!25}{white}
	\begin{tabular}{|p{45pt}|r|r|r|r|r|r|r|r|r|r|}
		\hline
		\rowcolor{gray!25}
		\multirow{2}{*}{}       & \multicolumn{2}{c|}{\caffe}            & \multicolumn{2}{c|}{\keras}            & \multicolumn{2}{c|}{\tensor}           & \multicolumn{2}{c|}{\theano}           & \multicolumn{2}{c|}{\torch}            \\ \cline{2-11} 
		& \multicolumn{1}{l|}{SO} & \multicolumn{1}{l|}{GH} & \multicolumn{1}{l|}{SO} & \multicolumn{1}{l|}{GH} & \multicolumn{1}{l|}{SO} & \multicolumn{1}{l|}{GH} & \multicolumn{1}{l|}{SO} & \multicolumn{1}{l|}{GH} & \multicolumn{1}{l|}{SO} & \multicolumn{1}{l|}{GH} \\ \hline
		Loss function           & 11.1\%                  & 0.0\%                       & 6.3\%                   & 1.2\%                       & 4.2\%                   & 7.8\%                       & 13.3\%                  & 6.25\%                      & 5.9\%                   & 4.9\%                       \\ \hline
		Network \linebreak connection        & 14.8\%                  & 11.8\%                      & 18.9\%                  & 10.2\%                      & 22\%                  & 13.3\%                      & 0.0\%                   & 21.9\%                      & 0.0\%                   & 26.8\%                      \\ \hline
		Add layer               & 11.1\%                  & 11.8\%                      & 5.6\%                   & 9.6\%                       & 2.5\%                   & 1.1\%                       & 0.0\%                   & 0.0\%                       & 5.9\%                   & 2.44\%                      \\ \hline
		Layer \linebreak dimension         & 3.7\%                   & 0.0\%                       & 7.0\%                   & 26.3\%                      & 13.6\%                  & 3.3\%                       & 13.3\%                  & 9.4\%                       & 11.8\%                  & 9.8\%                       \\ \hline
		Data dimension          & 22.2\%                  & 0.0\%                       & 22.4\%                  & 9.6\%                       & 11.9\%                  & 2.2\%                       & 26.7\%                  & 15.6\%                      & 23.5\%                  & 7.3\%                       \\ \hline
		Accuracy \linebreak metric         & 0.0\%                   & 0.0\%                       & 1.4\%                   & 0.0\%                       & 0.0\%                   & 0.0\%                       & 0.0\%                   & 0.0\%                       & 0.0\%                   & 0.0\%                       \\ \hline
		Data type        & 3.7\%                   & 29.4\%                      & 7.7\%                   & 13.8\%                      & 19.5\%                  & 10.0\%                      & 26.7\%                  & 6.2\%                       & 29.4\%                  & 14.6\%                      \\ \hline
		Activation       & 7.4\%                   & 0.0\%                       & 3.5\%                   & 3.6\%                       & 0.8\%                   & 0.0\%                       & 6.7\%                   & 12.5\%                      & 0.0\%                   & 2.4\%                       \\ \hline
		Iterations        & 7.4\%                   & 5.9\%                       & 4.95\%                    & 3.6\%                       & 5.9\%                   & 4.4\%                       & 0.0\%                   & 9.4\%                       & 5.9\%                   & 2.4\%                       \\ \hline
		Versioning           & 0.0\%                   & 0.0\%                       & 6.3\%                   & 9.0\%                       & 8.5\%                  & 51.1\%                      & 6.7\%                  & 0.0\%                       & 5.9\%                   & 0.0\%                       \\ \hline
		API contract            & 3.7\%                   & 0.0\%                       & 2.1\%                   & 1.2\%                       & 5.1\%                   & 1.1\%                       & 0.0\%                   & 3.1\%                       & 0.0\%                   & 0.0\%                       \\ \hline
		Data wrangling          & 0.0\%                   & 35.3\%                      & 4.2\%                   & 2.4\%                       & 1.7\%                   & 1.1\%                       & 0.0\%                   & 6.2\%                       & 0.0\%                   & 19.5\%                      \\ \hline
		Monitor & 11.1\%                  & 5.9\%                       & 2.8\%                   & 1.2\%                       & 1.7\%                   & 4.4\%                       & 0.0\%                   & 0.0\%                       & 0.0\%                   & 4.9\%                       \\ \hline
		Optimizer        & 0.0\%                   & 0.0\%                       & 1.4\%                   & 0.0\%                       & 0.0\%                   & 0.0\%                       & 0.0\%                   & 0.0\%                       & 0.0\%                   & 2.4\%                       \\ \hline
		Change neural arch.      & 3.7\%                   & 0.0\%                       & 5.6\%                   & 8.4\%                       & 2.5\%                   & 0.0\%                       & 6.7\%                   & 9.4\%                       & 11.8\%                  & 2.4\%                       \\ \hline
	\end{tabular}
\end{table}
\subsection{Data Dimension}
\label{subsec:data-dimension}

\finding{Fixing data dimension is the most common bug fix pattern (18.8\%) in
	\sof that can affect the robustness of DNN model.}
A large number of bugs (59 out of 415) in \sof are fixed by changing the data
dimension. 
This suggests that most DNN
models can easily be broken if the data processing pipeline changes or a
different format of data is fed to the DNN.
For example, in the following code snippet, we see how the bug
discussed in a
\sof post%
\footnote{https://stackoverflow.com/questions/37666887}
is fixed by adding a dimension to the input images.
\begin{lstlisting}[language = Pythonna]
model = Sequential()
...
model.compile()
model.load_weights('./ddx_weights.h5')
img = cv2.imread('car.jpeg', -1) # this is is a 32x32 RGB image
img = np.array(img)
+ img = img.reshape((1, 3, 32, 32))
y_pred = model.predict_classes(img, 1)
print(y_pred)
\end{lstlisting}
In the listing, the developer wants to read a CIFAR-10 image whose
dimension is (32,32,3) but the expected image size was (1,3,32,32).
Data dimension change can be categorized into the following kinds.
\paragraph{Resize} Resizing the input data is common, e.g. resizing an
input image of shape \code{(190,150)} to \code{(150, 150)}. A risk in this kind
of fix is the loss of information from the input due to resizing. Surprisingly,
this risk is never stated in the fixes presented on the bug fixes we have
studied. 
11 out of the 59 data dimension fixes involve resizing the data. Resizing can be
done in two ways: downscale or upscale. The downscale is the method where the
risk due to data loss is critical from our observation. Upsampling does not have
this risk of data loss, and recent results suggest that adding noise to the data
can potentially increase the robustness of a model \cite{xie2017mitigating}.

\finding{63\% of the resize related posts in \sof utilize the downscaling that
	can decrease the robustness of a DNN.}

7 out of the 11 data resizing post in \sof involves downscaling.
Downscaling decreases the robustness and \cite{xiao2017wolf} 
has shown that a simple resize downscaling operation can have a negative impact
on the
robustness. During downscaling, significant information loss occurs, and that
eventually decreases the features learned by the DNN. A DNN trained with
downscaled images will be easier to attack compared to the one trained with
original images. Our findings suggest that it would be useful to verify
the effect of the resizing fix on the vulnerability of the DNN.


\paragraph{Reshape} Reshaping the input occurs when the input vector
shape is changed. For example, a vector of size \code{{(32, 32)}} is changed
to \code{(1,32,32)}. In this case, no data loss happens and the tensor order is
changed from 2D to 3D. An example of this fix is presented in the \sof post
\#41563720%
\footnote{https://stackoverflow.com/questions/41563720}.
The reshaping does not lead to data loss. 
38 out of 59 data dimension fixes involve reshaping the dimension of the input.
Reshape may also involve changing the dimension through one hot encoding like
the following code snippet to fix \sof post \#49392972%
\footnote{https://stackoverflow.com/questions/49392972}:
\begin{lstlisting}[language = Pythonna]
train_labels = to_categorical(train_labels)
\end{lstlisting}


\paragraph{Reorder} To make this kind of dimension change, the input
data is ordered mostly to change the channel position. 
In image classification problems, channel refers to the color channels of three
primary colors.
(height, width, channel) represents the typical structure of a 3D image.
For example, the input of
shape \code{(32,32,3)} is changed to \code{(3,32,32)} to fix some bugs. Here the
channel number is moved to the first argument from the third argument. It can
also involve changing the image dimension order format like from \code{RGB} to
\code{BGR} as in the following snippet for fixing \sof post \# 33828582%
\footnote{https://stackoverflow.com/questions/33828582}:
\begin{lstlisting}[language = Pythonna]
img = caffe.io.load_image( "ak.png" )
+ img = img[:,:,::-1]*255.0 # convert RGB->BGR
\end{lstlisting}
\finding{Reorder and reshaping (79.7\% of the data dimension fixes in \sof) need an understanding of the specifications of the DNN layers as well as the libraries.}
9 out of 59 data dimension fixes involve reordering the dimension of inputs.
This is done because some of the libraries require dimension in a specific
order. These fixes are seen in the bugs where the developer works with multiple libraries having different channel position requirements in the image data, such as \sof post \#45645276%
\footnote{https://stackoverflow.com/questions/45645276/}.
DNN training can be assumed as a gradient descent based optimization problem, which can be computed when all the functions utilized in the model creation are differentiable. Data should be changed in such a fashion that does not affect the gradient descent computation to avoid side effects. In \emph{reshape} and \emph{reorder}, the only changes occur is the addition of dimension and reordering of the values that do not impact the gradient descent computation. So these changes theoretically have no side effects in the DNN models' behavior. 


\subsection{Layer Dimension}
\label{subsec:layerdim}
\finding{In \gh layer dimensions fixes are used more frequently (15.6\%) to fix
	the crash related bugs (75.9\%).}
In \gh, data dimension related fixes involve 7.5\% of all the fixes.
On the other hand, fixing the layer dimensions to make the DNN compatible with input data is a more common practice in \gh. 
Dimension related fixes can be done by analyzing the input and output of the
	layers by converting a neural network into a data flow graph. This kind of fixes includes dimension reduction or
	addition based on the adjacent layers' structure. However, these fixes can be either done by changing the data dimension to match the data with the layer dimension or vice-versa. The choice of the fix has an impact on the performance of the model. This phenomenon is known as the \emph{curse of
		dimensionality \cite{friedman1997bias}}. The curse of dimensionality states that increasing or decreasing the dimension can lead to overfitting/underfitting problems.
	PCA \cite{jolliffe2011principal}, T-SNE \cite{maaten2008visualizing} are some examples of the dimension reduction techniques that reduce the dimension of the features but these techniques suffer from the curse of dimensionality. To build an automated approach to avoid this side effect, a tool needs to optimize the performance of the model by either changing the data dimension or the layer dimension. AutoML
	\cite{jin2018auto} has done some preliminary work in this field that restructures
	the model by changing the layer dimension and adding layers to
	increase the performance. To the best of our knowledge, no tool currently exists that analyzes both data dimension and layer dimension changes to
	pick the optimum operations for a DNN model.

\subsection{Version-related Fixes}

\finding{Versioning-related bug fixes are the highest (17.6\%) in \gh indicating the high maintenance cost in DNN software due to library versioning.}
	
We have found that in \gh, long-running projects have to fix a lot of bugs due to frequently changing versions of the DNN libraries. 
	A number of these fixes require changing the API signatures
	to match with changes in the libraries.
	We have also observed a more complicated fix pattern for projects that use \tensor library as discussed in \S \ref{subsec:release}.
	\tensor often makes invasive, backward-incompatible changes adding difficulties to fix the introduced bugs. This indicates that the maintenance cost in DNN software is high.

%

\subsection{Network Connection}
\label{subsec:graphcon}
\finding{Network Connection is a prevalent fix in both \sof (17.8\%) and \gh
	(14.1\%) to fix crash (57.14\%), incorrect functionality (16.19\%),
	and bad performance (12.38\%) effects.}

The tensor and data flow through the network in a DNN  
during forward and backward propagation or prediction. 
For a smooth flow of data, the end-to-end connectivity in the network is essential. 
57 out of 415 fixes require fixing or adjusting the connectivity in the network.
We have found three kinds of network connection repairs. 

\paragraph{Merge layers}
A number of repair cases fixed bugs by merging two parallel layers
into a single layer. For example, the following code snippet shows a fix,
\begin{lstlisting}[language = Pythonna , basicstyle = \tiny]
+ main_branch.add(Merge([branch_1, branch_2], mode = 'dot'))
\end{lstlisting}
where two different branches are connected through dot product. 
The network was disconnected in the bug leading to a crash. 
%

\paragraph{Add feedback loops and input layers}
In some bug fixes, a feedback loop is added in the DNN model. 
In some of the fixes, the model is connected to the input via an input layer
like the following:
\begin{lstlisting}[language = Pythonna, basicstyle = \tiny]
+ lstm_out = LSTM(128, input_shape=(maxlen, len(chars)))(net_input)
\end{lstlisting}

\paragraph{Transfer learning}
Transfer learning is a popular technique that takes an already-trained network
with a different dataset. Then, the new model modifies the last layers to support
the goal of the new problem and then performs some retraining without modifying
the weights and biases of the layers from the previous network.
We have observed several network connection fixes needed when the developer is 
attempting transfer learning.
Generally these fixes change the last few layers of the DNN.
One such kind of fix is shown below from \sof post \#57248121\footnote{https://stackoverflow.com/questions/57248121/}:
\begin{lstlisting}[language = Pythonna]
+ model_final.fit_generator(train_generator.flow(np.array(X_train), np.array(y_train), batch_size=32),
+ validation_data=test_generator.flow(np.array(X_test), np.array(y_test), batch_size=32),
+ steps_per_epoch=len(X_train)/32, 
+ validation_steps=len(X_test)/32,
+ epochs=50)
\end{lstlisting}
In this example, the developer wants to train the imagenet with a pretrained network \code{VGG19} that has been used for face recognization. In this bug, the developer does not provide the correct data input size that leads to an error and fix was to include a data generator that loads the training data as expected by the \code{VGG19} model.

\subsection{Add Layer}
\finding{30\% of the add layers related fixes in \sof includes adding \code{Dropout} layer that can increase the training time $\sim$2-3x. }
In a DNN model, adding layers helps to increase the performance and learn the features more accurately. We have found that a vast majority ($\sim$30\%) of these bug fix patterns
includes the addition of the dropout layer. Dropout layer helps in removing the
effect of the overfitting \cite{srivastava2014dropout}  that can also be
achieved by using backpropagation. According to \cite{srivastava2014dropout},
backpropagation works better for the training dataset but does not work for new
data. Dropout layer randomizes the structure of the architecture that helps the
neural network to learn more features with every iteration. Dropout layer
removes the connection between layers randomly stopping the data flow through those nodes and edges. Randomly reducing connections can have a negative impact on training time. With Dropout layers, the convergence of the training takes $\sim$2-3x more time \cite{srivastava2014dropout}. 

\subsection{Loss Function}
\finding{Among DNN hyperparameters, change of loss function happens to fix 6.2\%
	(highest) of the bugs in \sof and 3.7\% in \gh that helps to enhance prediction accuracy and increase the
	robustness against adversarial attacks.}
Loss function plays a crucial role in the convergence of the training process and in getting better accuracy during prediction. 
A model with wrong loss function does not learn the decision boundary of the features well and there can be overlap between the decision boundaries \cite{janocha2017loss, pan2019identifying} in the high dimensional feature space making the model vulnerable to adversarial attacks \cite{saito2018effects}. 
By a careful and deeper analysis of these loss function related fixes, we have found that they can be categorized into the following kinds:

{\em Add new loss function.\ } The fixes in this category involve
adding a custom or built-in loss function.
10 out of 23 fixes fall into this category. In some of the fixes, it is needed to
modify the network connectivity for the new loss function to work.  For example, in
the following fix of the bug in \sof post  \#51257037\footnote{https://stackoverflow.com/questions/51257037/}, the last layer
is kept outside the gradient descent computation during training by adding \code{trainable = False}. 
\begin{lstlisting}
output = Dense(1, trainable = False)(hidden_a)
\end{lstlisting}

The custom loss function was designed by the developer in such a way that all 
but the output layer participate to lead to the convergence of the model.  
However, the convergence was not successful,  as the output layer was actively 
participating in the forward and backward propagation that caused an abrupt 
change in the value of the loss function.  
Fixing these bugs require live trainable parameter analysis. 
This approach will help to monitor the active trainable parameters during the 
training to localize and fix these bugs. 
Currently, the developer needs to rely on theoretical knowledge to fix these 
bugs due to the lack of such kind of analysis frameworks. 

{\em Change loss function.\ } 9 instances of bug fixes fall into the
category of changing the loss function. 
Our analysis of these fixes reveals that the choice of these loss functions is sometimes confusing. Developers need to understand the data properties and the goal of the DNN task to come up
with a proper loss function. For example, in constructing DNN models for classification problems, the developers are confused between the choice of \code{binary_crossentropy} and
\code{categorical_crossentropy} as discussed in the fix of \sof post \#45799474\footnote{https://stackoverflow.com/questions/45799474/} and \sof
post \#42081257\footnote{https://stackoverflow.com/questions/42081257}.
The first loss function works better for the binary classification problems;
however, when the classification problem has more than two categories, one
should use \code{categorical_crossentropy} as a loss function to avoid poor performance. 
Sometimes, the fix involves adding some filter to the mathematical
operation used in the loss function. For example, we see the following bug fix of \sof post \#34223315\footnote{https://stackoverflow.com/questions/34223315/}
\begin{lstlisting}[language = Pythonna]
cross_entropy = -tf.reduce_sum(y_ * tf.log(tf.clip_by_value(y_conv, 1e-10,1.0)))
\end{lstlisting}
caused by the following line: 
\begin{lstlisting}[language = Pythonna]
cross_entropy = -tf.reduce_sum(y_ * tf.log(y_conv))
\end{lstlisting}
In the above code snippet, the problem is that the user will get \code{NaN} values if
\code{y_conv} becomes negative as the \code{log} of negative numbers is
undefined. 
The fix adds a clipper function to filter out negative values to the \code{log} operation. 
In another fix of the same kind of bug in \sof post \#42521400\footnote{https://stackoverflow.com/questions/42521400/}, \code{softmax} is
used as the filtering operation that stops propagating values $ <= 0$ to the \code{log} operation. 
\begin{lstlisting}
softmax = tf.nn.softmax(logits)
xent = -tf.reduce_sum(labels * tf.log(softmax), 1)
\end{lstlisting}

\subsection{Commonality of Fix Patterns in \sof and \gh}
\finding{The p-value is 0.83 between the bug fix pattern distributions of \sof and \gh indicating commonality of bug fix patterns in \sof and \gh. }
We have conducted a t-test at 95\% significance level to understand the 
distribution of bug fix patterns in \sof and \gh. 
The null hypothesis is \emph{$H_0: $ the distributions are the same}. 
The null hypothesis is to be rejected if the p-value is less than 5\% or 0.05. 
Our computation shows that the p-value is very high (0.83). 
So, $H_0$ can not be rejected concluding that the distributions are similar. 
We also notice that though in some bug fix categories e.g., data dimension, 
layer dimension, and versioning, there is a significant difference among the 
\sof and \gh distributions, the other categories have a similar share of 
occurrences in \sof and \gh. 
This indicates that the bug fix patterns have commonality across \sof and \gh.

\section{Fix patterns across bug types}

\begin{figure}
	\includegraphics[keepaspectratio = True, scale = .4, trim={0cm 5cm 0cm 6cm},clip]{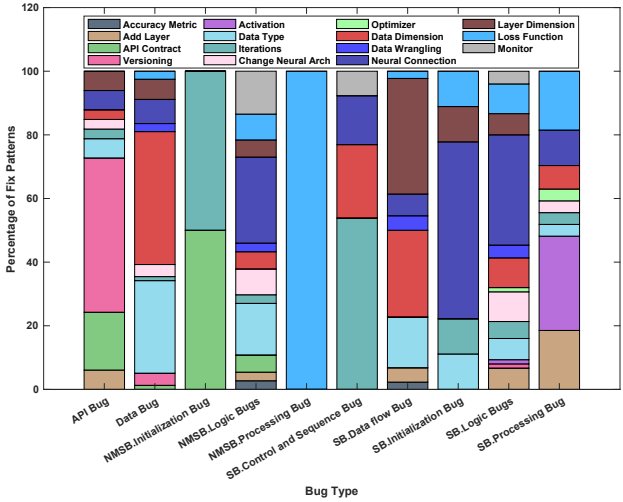}
	\caption{Distribution of Bug Fix Patterns for Different Bug Types \sof}
	\label{fig:bugtypepatternSO}
\end{figure}
\begin{figure}
	\includegraphics[keepaspectratio = True, scale = .4, trim={0cm 5cm 0cm 6cm},clip]{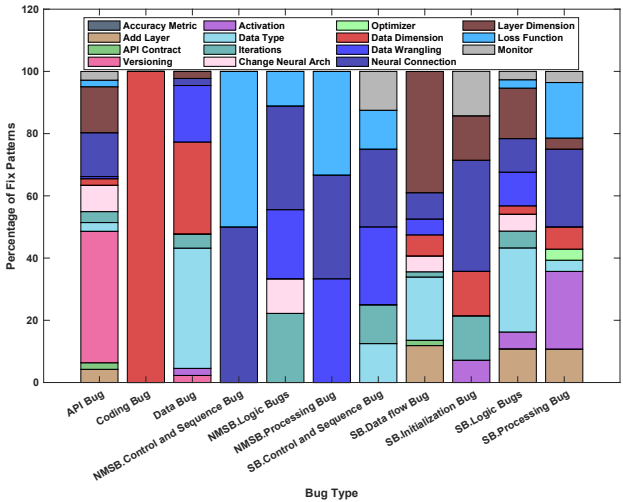}
	\caption{Distribution of Bug Fix Patterns for Different Bug Types \gh}
	\label{fig:bugtypepatternGit}
\end{figure}

To answer RQ2, we analyze the correlation between the bug types in the bug dataset presented by
\cite{islam2019comprehensive} and the bug fix patterns studied by this paper using the distribution of the bugs and their corresponding fixes.
The distribution of bug fix patterns across different bug types  in \sof and \gh are shown
in the \figref{fig:bugtypepatternSO} and \ref{fig:bugtypepatternGit}, respectively. The
horizontal and the vertical axes describe the different bug types from
\cite{islam2019comprehensive} and the percentage of different fix patterns
needed to fix those bugs, respectively. 
\finding{For API bugs, fixing of the specifications between APIs is dominant (42\% in \sof and 48\% in \gh ) .}
Fixing API specifications involves changing API contracts due to API versioning and supporting
inter-library operations within a model. 
Fixing API specifications is needed due to the following reasons:

{\em Change of specifications due to version upgrade.\ } 20 fixes in \sof involve changing specifications which are required due to the change of the library
version. 
The changes during the upgrade of the library version involves the following changes:
change fully qualified method names, change API signature, and
change probabilistic behavior of the APIs.
Though fixes due to the change of fully qualified method names and change of API signature are well-studied problems \cite{dietrich2014broken, brito2018and, jezek2015java}, the fixes due to the change of probabilistic behavior of the APIs are hard and different from traditional API changes. Localizing of these bugs are difficult due to the lack of sophisticated probabilistic analysis tools for DNN. For example, the bug discussed in \sof \#49742061\footnote{https://stackoverflow.com/questions/49742061/} says that the results
are different in two versions of \tensor. The fix of this bug involves adding a
dead code line that tweaks around the
underlying probabilistic behavior of the APIs  by overriding the modified random seed. 
\begin{figure}
	\includegraphics[scale = .4, width=0.7\linewidth]{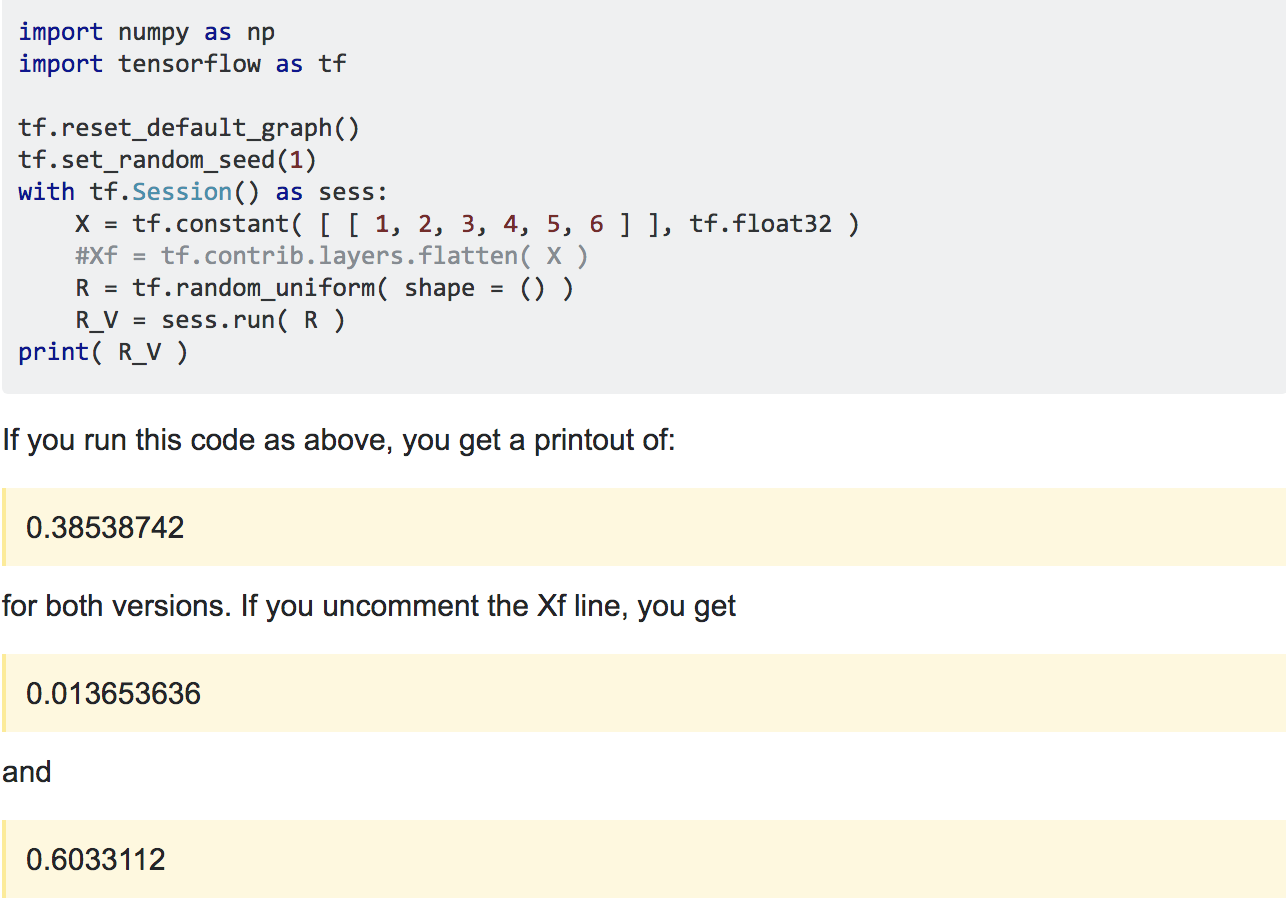}
	\caption{Fix of \sof \#49742061}
	\label{fig:so1}
\end{figure}
The fix of \sof \#49742061 is shown in \figref{fig:so1}. The fix adds the line \code{Xf =
	tf.contrib.layers.flatten( X )} before the line \code{R = tf.random_uniform(
	shape = () )}. This addition overrides the random seed in the new version with
the one in the previous version. 

\begin{figure}
	\includegraphics[scale = .35, width=0.8\linewidth]{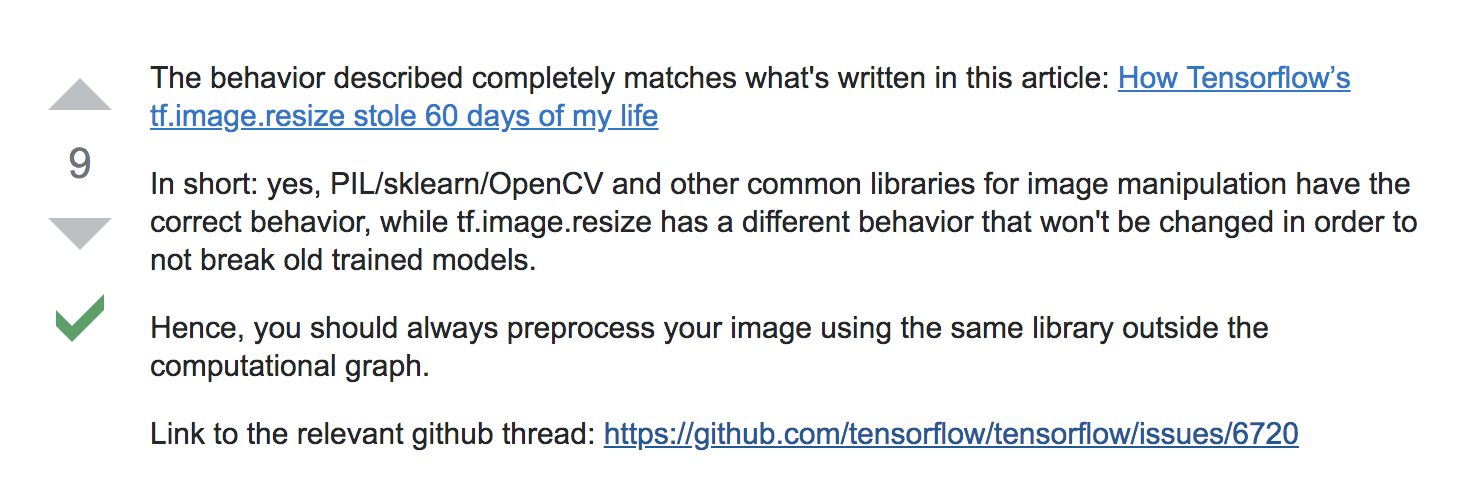}
	\caption{Fix of \sof \#54497130}
	\label{fig:so2}
\end{figure}
Our observation gives the intuition that the fix of versioning bugs due to the change of the probabilistic distribution in different version needs new DNN specific probabilistic analysis techniques.

{\em Change specification to support interlibrary.\ } In these
fixes, the DNN program uses more than one library. These bugs arise due to
the similar assumption of the behavior and specifications for different APIs in different  libraries. Fixing
of these bugs requires the expertise in both the libraries e.g., the bug discussed in \sof \#54497130\footnote{https://stackoverflow.com/questions/54497130/} that is shown in \figref{fig:so2}.
The discussion points to an issue in the official \tensor repository. The
solution suggested to avoid using APIs from other libraries to pre-process
images. 
However, in similar scenarios, the use of specialized image processing libraries is recommended to get better performance. 


From \figref{fig:bugtypepatternSO} and \ref{fig:bugtypepatternGit}, we have found that fixing the data dimension is the most prominent pattern (41.77\%) for fixing data
	bugs in \sof. For fixing data bugs in \gh, the most prominent fix patterns are the change of data type (38.64\%) and data dimension (29.55\%).
	This suggests that for fixing data bugs, the major changes are related to data dimensions. This happens because the dimension of the data is very important for the correct functionality of the DNN model. 

 For fixing logic bugs the most common practice is to modify the network
connectivity ($\sim$27.03\% in \sof and $\sim$33.33\% in \gh). A detailed discussion on network connectivity  is presented in \S \ref{subsec:graphcon}.
Whereas, a significant amount of data flow bugs can be fixed by changing the layer
dimension ($\sim$36.36\% in \sof and $\sim$38.98\% in \gh). A detailed discussion on fixing layer dimension is presented in \S \ref{subsec:layerdim}.

These observations give us the intuition that for fixing different types of bugs,
unique technical approaches might be needed.


\section{Fix Patterns across libraries}

To answer RQ3, we have studied the distribution of fix
patterns across the 5 libraries. Then, we have conducted statistical pairwise t-test at 95\%
significance level between the libraries. 
Table \ref{tbl:ttest} shows the p-values
found from this test across the libraries.

\begin{table}[h]
    \footnotesize
	\centering
	\caption{P-value of the distribution of Bugs btween the libraries}
	\rowcolors{2}{gray!25}{white}
	\begin{tabular}{|  l  |  r |  r |  r |  r | r |}
		\hline
	\rowcolor{gray!50}
    Library & \caffe & \keras & \tensor & \theano & \torch \\
    \hline
    \caffe & 1.0 & 0.0045 & 0.00735 & 0.19 & 0.30\\
    \hline
    \keras & 0.0045 & 1.0  & 0.84 & 0.0021 & 0.0024 \\
    \hline
    \tensor & 0.0073 & 0.84  & 1.0 & 0.0039 & 0.0044 \\
    \hline
    \theano & 0.19 & 0.0021  & 0.0039 & 1.0 & 0.80 \\
    \hline
    \torch & 0.30 & 0.0024  & 0.0044 & 0.80 & 1.0 \\
    \hline
\end{tabular}
	\label{tbl:ttest}
\end{table}

We assume the null hypothesis is 
\emph{$H_0$: the distribution of the fix patterns across two libraries are
	same. }
If the p-value is less than 5\% or 0.05, then we reject $H_0$. The
p-value for the library pairs \caffe-\theano (.19), \caffe-\torch (.30),
\keras-\tensor (0.84), \theano-\torch (0.8) are much greater than 5\%. 
So in these, cases we can not reject the null hypothesis. 
So, the libraries \caffe, \theano, and \torch show similar kind of bug fix patterns. 
The pair \keras-\tensor form a very strong related group with a p-value close to 100\%. 
This suggests that similar kinds of automatic bug fix tools may be reused 
for \caffe, \theano, and \torch after converting into a common intermediate 
representation. 
Similarly, \keras and \tensor bugs can be fixed using similar technical approaches.

\section{Introduction of Bugs Through Fixes}
\label{sec:fixnew}
\finding{29\% of the bug fixes introduce new bugs in the code adding technical debt \cite{sculley2014machine} and maintenance costs.}
To answer RQ4, we have analyzed 100 randomly chosen fixes from \sof to understand
whether fixing a bug can introduce a new bug. We have read the replies to the answers selected by filtering criteria discussed in \S \ref{sec:methodology}.
Then, we have identified whether the fix introduced new bugs by analyzing all replies to the answer fixing the original bug and classify them into bug type, root cause, and
impact using the classification scheme proposed by the prior work
\cite{islam2019comprehensive}. We have found that 29\% fixes in the randomly
sampled dataset introduce at least one new bug in the code. 
Here, a new bug indicates that the original bug was fixed by the solution posted; 
however, the solution introduces a  new bug that is different from the original bug type. 
Furthermore, we have compared the bug
type, root cause, and the effect of the bugs of \sof posts with the newly
introduced bugs and have found that only 6.8\%, 13.8\%, and 24.1\% of the bugs match
the classification of the parent bug type, root cause, and impact, respectively.
This result shows that a majority of the bugs introduced are of new types and
their behavior is entirely different than that of the parent bugs'. 
In the Table \ref{tbl:rq5}, we have shown the distribution of the new bugs across 
the different libraries and how these new bugs are classified into different 
categories of bug type, root cause, and impact. 
We have also found that the \emph{Crash} (55.8\%) is the most common
impact of these new bugs and a majority of these bugs are of \emph{API Bug}
(37.9\%), and the most common root cause of these bugs are \emph{API Change}
(34.5\%) that includes the change of either the signature of the API 
or the fully qualified name of the API.
44.8\% and 34.5\% of the newly introduced bugs are from
\keras and \tensor. \caffe, \theano, and \torch related bug fixes introduce 10.34\%, 3.45\%, and 6.90\% new bugs, respectively.
\begin{figure*}
	\centering
	\null\centerhfill
	\subfloat[Root Cause]{%
		
		\includegraphics[trim={5.5cm 3cm 3cm 2cm},width=0.135\textwidth]{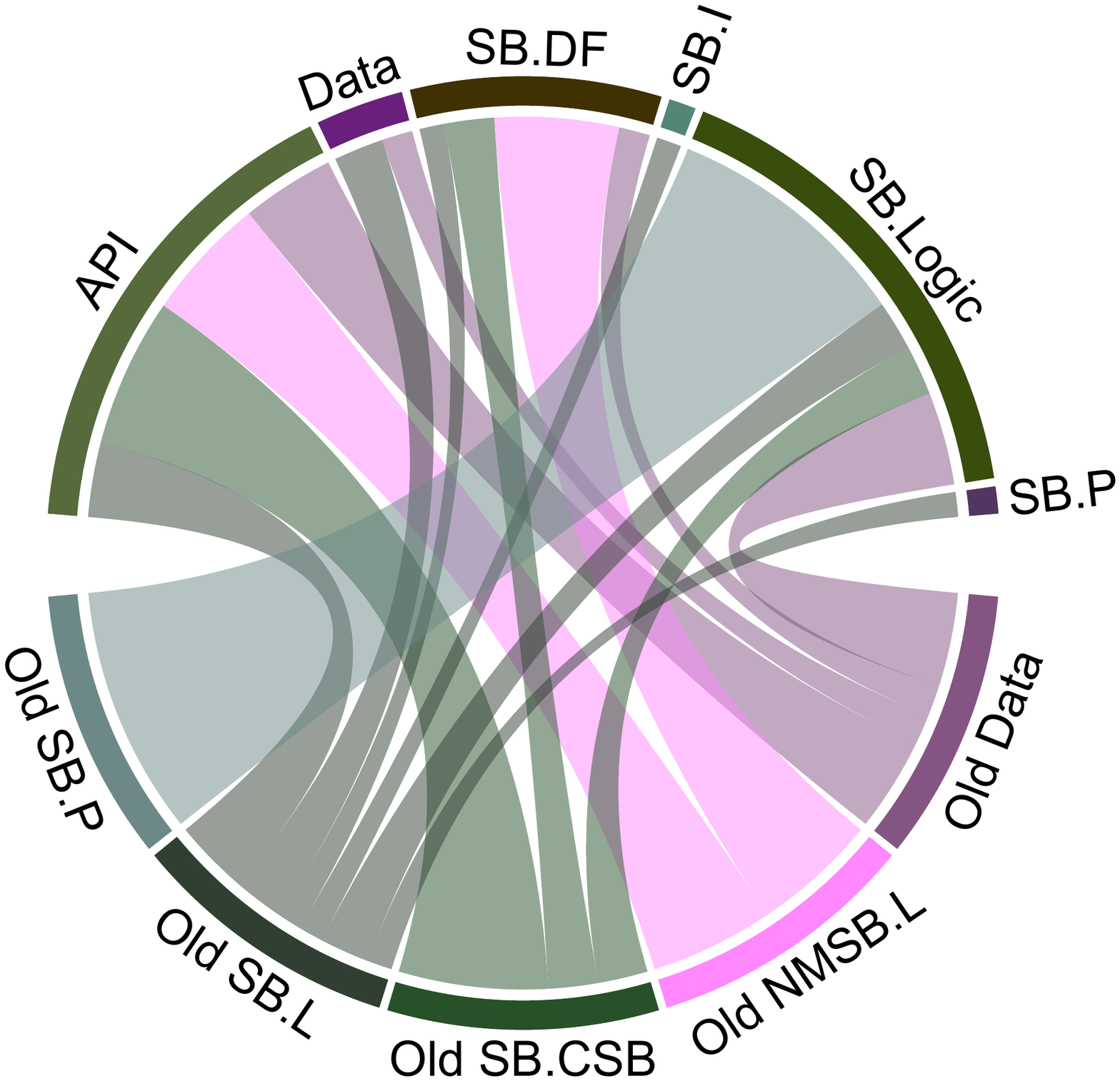}
		\centering
	}
	\hfill
	\subfloat[Bug Type]{%
		\includegraphics[trim={5.5cm 3cm 3cm 2cm},width=0.135\textwidth]{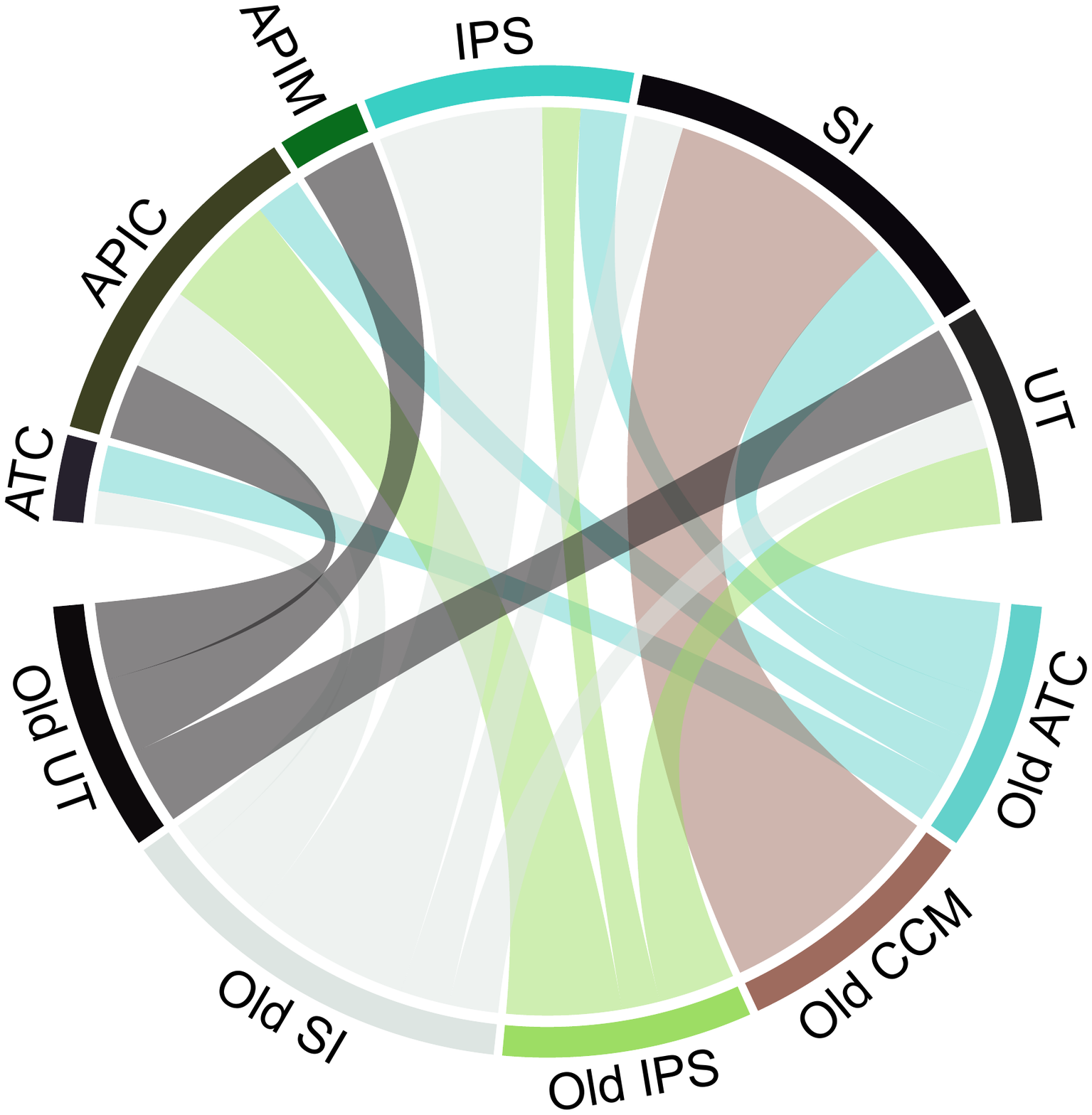}
		\centering
	}
	\centerhfill
	\subfloat[Impact]{%
		\includegraphics[trim={5.5cm 3cm 3cm 2cm},width=0.135\textwidth]{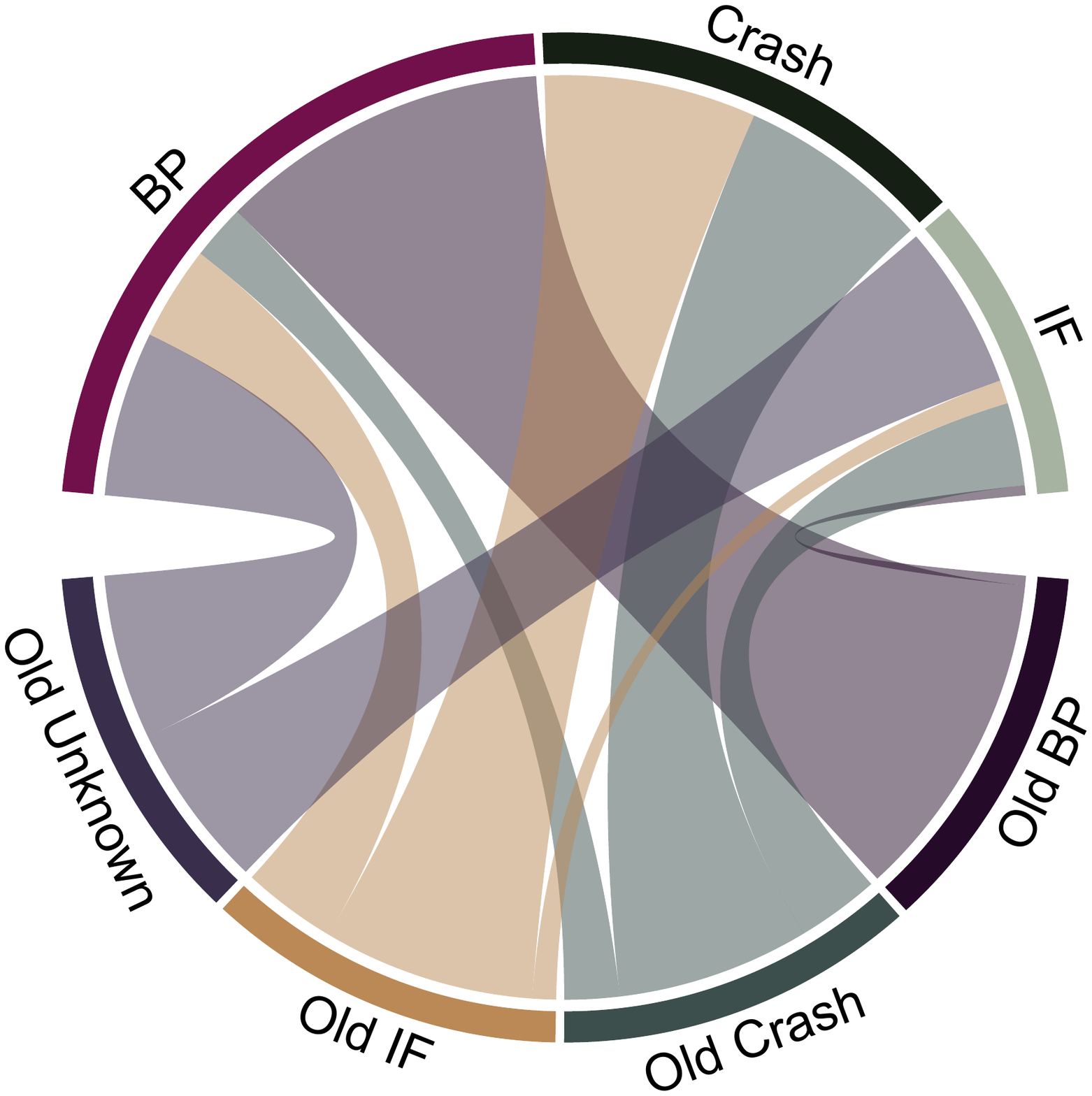}
		\centering
	}
	\centerhfill
	\caption{Bug fix pattern distribution: SB.P$\rightarrow$SB.Processing,
		SB.L$\rightarrow$SB.Logic, DF$\rightarrow$Data
		Flow,SB.I$\rightarrow$SB.Initialization, ATC$\rightarrow$Absence of Type
		Checking, BP$\rightarrow$Bad Performance, IF$\rightarrow$Incorrect Functionality
	}
	\label{fig:rq5}
	\vspace{2em}
\end{figure*}

\begin{table*}
	\tiny

\centering
\caption{Statistics of the Introduction of New Bugs During Bug Fix}
\begin{tabular}{|l|l|l|l|l|l|l|l|l|l|l|l|l|l|l|l|}

\hline
\rowcolor[HTML]{C0C0C0} 
\cellcolor[HTML]{C0C0C0}                          & \multicolumn{6}{c|}{\cellcolor[HTML]{C0C0C0}Bug Type} & \multicolumn{6}{c|}{\cellcolor[HTML]{C0C0C0}Root Cause} & \multicolumn{3}{c|}{\cellcolor[HTML]{C0C0C0}Impact} \\ \cline{2-16} 
\rowcolor[HTML]{C0C0C0} 
\multirow{-2}{*}{\cellcolor[HTML]{C0C0C0}Library} & API Bug & Data Bug & SB.DF  & SB.I & SB.L    & SB.P   & ATC     & APIC    & APIM  & IPS     & SI      & UT      & Bad performance       & Crash        & IF           \\ \hline
\caffe                                             & 0\%     & 0\%      & 0\%    & 0\%  & 100.0\% & 0\%    & 0\%     & 0\%     & 0\%   & 33.3\%  & 66.7\%  & 0\%     & 66.7\%                & 0\%          & 33.3\%       \\ \hline
\keras                                             & 30.8\%  & 7.69\%   & 30.7\% & 0\%  & 23.1\%  & 7.69\% & 7.69\%  & 30.8\%  & 0\%   & 15.4\%  & 15.4\%  & 30.8\%  & 23.1\%                & 61.5\%       & 15.4\%       \\ \hline
\tensor                                                & 60.0\%  & 20.0\%   & 0\%    & 10\% & 10.0\%  & 0\%    & 20\%    & 60\%    & 0\%   & 10\%    & 10\%    & 0\%     & 20\%                  & 70\%         & 10\%         \\ \hline
\theano                                                & 0\%     & 0\%      & 0\%    & 0\%  & 100\%   & 0\%    & 0\%     & 0\%     & 0\%   & 0\%     & 100\%   & 0\%     & 100\%                 & 0\%          & 0\%          \\ \hline
\torch                                                & 50.0\%  & 0\%      & 50\%   & 0\%  & 0\%     & 0\%    & 0\%     & 0\%     & 50\%  & 0\%     & 0\%     & 50\%    & 0\%                    & 50\%         & 50\%         \\ \hline
\end{tabular}
\label{tbl:rq5}
\end{table*}
\finding{37.9\% of the new bugs are from API Bug, 34.5\% of them are due to API
	Change, and 55.2\% of them end in a crash.}
In \figref{fig:rq5}, the relation between the parent bugs' root cause,
type and effect with the newly introduced bugs' distribution has been visualized.
In this visualization, the \emph{old} represents the parent bug and the relation
has been drawn by a connection between two bug distributions. The width of the
connection determines the strength of the relation. The perimeter covered by each
bug type/root cause/impact depicts its overall strength.
We have found that a large section of bug fixes introduces API
bug and the major reason for that is the API change that mostly due to the
versioning of the APIs and these fixes primarily lead to a crash and bad
performance.

\section{Challenges in Fixing Bugs}
\label{sec:challenge}
In this section, we explore the answer to RQ5 to identify the common challenges
faced by 
the developers in fixing the DNN bugs.
To understand the challenges, we have used a classification scheme
separate from bug fix patterns. 
Similar to the labeling performed for bug fix patterns, two 
raters have independently classified 
each post used in this study. These classes of new challenges are described
below:

\subsection{DNN Reuse}
\label{subsec:outsourced}
Training DNN models can be expensive because it requires sophisticated 
computational resources and a large amount of labeled data that might not be 
readily available. This has led to the reuse of DNN models that
creates unique issues such as backdoor attack~\cite{chen2017targeted},
injection of bias~\cite{biggio2012poisoning}, and mismatch
between the intent of the pretrained DNN model and the intent of the developer.

\begin{lstlisting}[language = Pythonna, basicstyle = \tiny]
base_model = ResNet50(input_shape=(224, 224, 3),
include_top=False,weights='imagenet',pooling='avg')
+ x=base_model.output
+ x = Dense(512, activation='relu')(x) #add new layer
+ x = Dropout(0.5)(x) #add new layer
+ x = Dense(512, activation='relu')(x) #add new layer
+ x = Dropout(0.5)(x)
\end{lstlisting} 
In the example above from \sof post \#
49226447\footnote{https://stackoverflow.com/questions/49226447},
the developer wants to train a predefined DNN model structure \code{ResNet50} 
using the cancer dataset. 
The trained network results in overfitting as the developer was not 
aware of the structure of the reused model and needed to modify the 
network by adding dropout and dense layers to reduce the 
effect of overfitting.

\subsection{Untraceable or Semi-Traceable Error}
\label{subsec:error}
In case of a crash bug, the common strategy to localize the bug is to analyze
the error message. 
However, we have found that bug localization is very challenging in DNN software
because errors and faults are non-trivially related. 
To illustrate, consider the code snippet below from 
\sof post \#
33474424\footnote{https://stackoverflow.com/questions/33474424}: 
\begin{lstlisting}[language = Pythonna, basicstyle = \tiny]
model = Sequential()
model.add(Dense(hidden_size, input_dim=input_size, init='uniform'))
model.add(Activation('tanh'))
...
y_pred = model.predict(X_nn)
\end{lstlisting} 
This code produces the following error trace:
\begin{lstlisting}[language = Pythonna, basicstyle = \tiny]
ttributeError                            Traceback (most recent call last)
<ipython-input-17-e6d32bc0d547> in <module>()
1 
----> 2 y_pred = model.predict(X_nn)
491     def predict(self, X, batch_size=128, verbose=0):
492         X = standardize_X(X)
--> 493         return self._predict_loop(self._predict, X, batch_size,
verbose)[0]
494 
495     def predict_proba(self, X, batch_size=128, verbose=1):

AttributeError: 'Sequential' object has no attribute '_predict'

\end{lstlisting} 
From this error message, a developer might start digging into 
the code of \code{predict} function and the \code{Sequential} object;
however, the issue is the missing compilation step. 
Due to this, the model
connection is not initialized and error propagates to the \code{predict}
operation and halts the training process. 
We have studied randomly 50 bugs yielding \code{Crash} from \sof. 
\textbf{We have found that 11 out of 50 posts does not indicate 
	any error message and in rest of the 39, 20 posts have 
	a fix that does not match with the error message.}

\subsection{Fast and Furious Releases}
\label{subsec:release}

We have previously discussed that a large number of fixes are due to the
rapid versioning and invasive changes in DNN libraries. 

\begin{table}[h]
	\vspace{5pt}
	\centering
	\footnotesize
	\caption{\tensor API changes. Change= \# of operations changed in comparison to the previous version.}
	\rowcolors{2}{gray!25}{white}
	\begin{tabular}{| l | r| r|r|}
		\hline
	\rowcolor{gray!50}
    Version & \# of Symbols & Change & Release Date \cite{tensor}\\
    \hline
    \hline
    v2.0 (Beta) & 6504 & 2185 &  Jun 7, 2019\\
    \hline
    v1.14.0 & 8363 & 59 & Jun 18, 2019 \\
    \hline
    v1.13.1 & 3560 & 39 & Feb 25, 2019 \\
    \hline
    v1.12.0 & 3314 & 52 & Nov 1, 2018\\
    \hline
    v1.11.0 & 3145 & 175 & Sep 25, 2018\\
    \hline
    v1.10.0 & 3230 &  N/A & Aug 7, 2018\\
    \hline
\end{tabular}
	\vspace{5pt}
	\label{tbl:api}
\end{table} 

To study this challenge, we have labeled all removed, reconfigured, 
or renamed operations of \tensor from version 1.10 to 2.0 (latest in June 2019).

In Table \ref{tbl:api}, we have shown the number of symbols of operations
available for each \tensor releases and the number of operations that have been
deleted, renamed, or reconfigured in comparison to the previous version. 
\textbf{We have found that from the v1.14 to v2.0 ~26\% of
	the operations have been changed}. 
We have also studied \keras v2.0, v2.1, v2.2, and v2.3 to understand whether 
this problem is only prevalent in \tensor or not. 
Our study has found that during the transition from v2.0-v2.1, v2.1-v.2.2, 
and v2.2-v2.3, the percentage of changes in operation are 6\%, 8\%,
and 4\%, respectively.

A non-trivial challenge for repairing DNN software is the probabilistic behavior
of the APIs.
Some of these version upgrades also change the probabilistic behavior
of the APIs causing some difficult bugs. An example is presented below where the
change of the 
probabilistic distribution changes the output of the same operation with
different
versions\footnote{https://stackoverflow.com/questions/49742061}.
\begin{lstlisting}[language = Pythonna, basicstyle = \tiny]
with Tensorflow 1.3
Z3 = [[-0.44670227 ...  0.46852064]
[-0.17601591 ...  0.5747785 ]]
with Tensorflow 1.4+
Z3 = [[ 1.44169843 ...  1.36546707]
[ 1.40708458 ...  1.26248586]]
\end{lstlisting}

\section{Threats to Validity}
\label{sec:tv}
{\em External Threat.\ }
A source of external threat can be the dataset used to study the bug repair
pattern. To alleviate this threat we use a benchmark dataset of DNN bugs
prepared by \cite{islam2019comprehensive}. 

{\em Internal Threat.\ }
An internal threat can be the coding approach used to classify the bug
fix patterns. We use the widely adopted open coding approach to come
with a classification scheme to minimize this threat. 
Two Ph.D. students independently came up with the classification schemes. 
Then, these schemes were merged through moderated discussions. 
The expertise of the raters can be another source of an internal threat. We
alleviate this threat by involving raters who have expertise in both the DNN
libraries and the bug fix patterns. The raters were also trained on the coding
scheme before the labeling. We also use kappa coefficient to measure the
inter-rater agreement throughout the labeling process. And the value of kappa
coefficient indicates that the labeling was successful with a perfect agreement.

Another threat is the number of posts in \sof for each library are not the same. 
To mitigate this threat, we have performed an ANOVA test on the \sof
bug fix patterns. 
We have found that F (0.0002) < F-critical (2.50) that implies
that the distribution of the bug fix in \sof is not unbalanced.

 \section{Discussion}
\label{sec:discuss}
We have analyzed the bug fix patterns in DNN and have found that
there are significantly different new patterns compared to the non-ML bug fix
patterns. 
There are also new challenges in fixing these bugs. 
In the analyses of RQ1, we have found that major fixes
are related to data dimension, layer dimension, network connection,
addition of layer, and loss function. 
To fix such issues, we
need to know the structure of the network, how data flowing through the network is
modified through various mathematical operations,
how the performance evolves during forward and backward propagation 
due to the use of loss function, accuracy metrics, etc. 
This presents a number of immediate research
challenges related to DNN API design and verification. 
To apply the data-related fixes, we need to understand the implicit
dependencies between the data and model. 
This problem is akin to the notion of implicit coupling between modules. 
Studying various
existing techniques to address strongly coupled data could be beneficial to fix the data-related problems.
To fix the connections among consecutive layers in a neural network,
the DNN model needs to be converted to a suitable common intermediate
representation (IR). Then, we need to perform a reachability analysis to find
the portion of the graph disconnected from the rest to fix such 
connection-related problems.
Also, the fixes related to the addition of layer and change of loss function can
be addressed
automatically by mining specifications related to such layers and loss function
from large codebases
\cite{uddin2012temporal, nguyen2014mining}. 


In RQ1, we have also shown that some of the bug fixes have
a negative impact on the robustness of DNN models \cite{carlini2017towards}.
Studying such cases further and developing tips for new developers is necessary 
so that they avoid falling into these traps without this knowledge.
In RQ2, we have seen that bug fixes are different for different bug types. 
We have noticed that fixing API bugs require fixing the specification between APIs. 
These fixes can be achieved by validating the compatibility among APIs by adding
robust test suites before releasing new versions.
In RQ3, we have identified the commonalities among the fix patterns of different
libraries.
Efforts on repairing bugs in DNNs are certainly needed, and they can focus on 
these commonalities to cover more ground quickly. 
In RQ4, we have observed that fixing bugs can lead to new bugs. 
Our findings identifies some common situations where this happens, e.g., fixing layer
dimension has the possibility of adding data bugs. 
We concluded our analyses by showing that new challenges are present in
fixing DNN bugs. Though some of these fixing strategies have been adopted by existing
tools, more work on validation and repair is warranted
in several SE sub-communities such as program 
analysis, runtime verification, formal methods, etc.
Analysis representation specific to DNNs can be developed to enable repair work.
Runtime monitoring framework for DNNs would be useful to prevent errors from 
occurring and to collect traces for dynamic analyses based repair techniques. 
Safety critical data science applications of DNNs need these efforts to become more 
dependable~\cite{RajanDependable2020}.

\section{Related Works}
\label{sec:related}

The closest related works are that by 
Zhang~\etal~\cite{zhang2018empirical}, Islam~\etal~\cite{islam2019comprehensive}
and Pan, Kim, and Whitehead~\cite{pan2009toward}.

{\em Study on traditional non-DNN Bugs.\ } 
Pan, Kim, and Whitehead~\cite{pan2009toward} have studied seven Java projects to
discuss the bug fix patterns in these projects. They have also proposed a
classification scheme in categorizing the bug fix patterns in Java. The
classification includes 9 broad categories and a total of 26 lower-level
categories. This prior research suggests that the IF-related and Method Call
(MC) related bugs are most frequent. In DNN bug fix strategies, the MC and
sequence addition or deletion related bug fix pattern is present. We do not find
any evidence of other bug fix strategies in DNN and that has inspired us to
derive a classification scheme using the open coding approach to
classify the DNN bug fix patterns.

Programming bugs are well-studied in software engineering.
There is a rich body of empirical studies on 
bugs, e.g.~\cite{saha2018bugs,
		ebert2015exploratory,ma2017developers,zaman2011security,lu2008learning,li2006have,chou2001empirical}
and bug repair, e.g. %
~\cite{park2012empirical,bachmann2010missing,zhong2015empirical,pan2009toward};
however, these works have not studied DNN bugs and repairs that have their own
set of 
unique challenges~\cite{islam2019comprehensive, zhang2018empirical,
islam2019developers, thung2012empirical}.

{\em Study on DNN Bugs.\ }
Zhang~\etal~\cite{zhang2018empirical} have studied bug patterns in \tensor using
both \gh and \sof. They have discussed the new patterns and characteristics of
the bugs by \tensor users to write DNN applications. They have also discussed
the three new challenges in detecting and localizing these bugs. The study was
limited to \tensor and also does not discuss the bug fix patterns. We generalize
to a number of deep learning libraries and identify the new patterns of fixing
the bugs in DNN software. We also discuss the new challenges in fixing these
bugs. We have discussed three new challenges in fixing DNN bugs.

Islam~\etal~\cite{islam2019comprehensive} have studied five different DNN
libraries. They have done a more general study on DNN bug characteristics beyond
\tensor. However, their work has not discussed the bug fix patterns and
challenges in fixing those bugs. Our work focuses on the identification of the
bug fix patterns and challenges in those fixes.
We have utilized the dataset prepared by this work.

Sun~\etal~\cite{sun2017empirical} studied the issues in 3 ML libraries \scikit,
\caffe, and \paddle to understand the bug fix patterns in these libraries.
However, we study the DNN models created using DNN libraries.
Our findings do not have any commonalities.

Zhang~\etal~\cite{zhang2019empirical} studied 715 \sof bug related posts for
TensorFlow, PyTorch, and Deeplearning4j to classify the questions into 7
different categories and built an automated tool that categorizes questions
based on the frequently found words from each category and computing the tf-idf
value with respect to the keywords. Also, the authors have studied the
challenges of answering the question in \sof by calculating the response time
for each category and have found 5 categories of root causes for the bug related
posts. Whereas, our study has been on the bug fixing strategies for 5 DNN
libraries.

Pham~\etal~\cite{pham2019cradle} studied three deep learning (DL)
libraries \tensor, \emph{CNTK}, and \theano to localize and detect deep learning
bugs using cross-validating different backend i.e., \tensor, \emph{CNTK}. In
contrast, our work studied five DL libraries, using bug-fix commits from \gh and
\sof posts, that allowed us to draw interlibrary observations of the fixing
strategies. Also, our work expanded the study to include deeper discussion about
each fix pattern, the common challenges developers face while fixing these bugs,
and how fixing bugs introduces a new bug.

\section{Conclusion and Future Work}
\label{sec:conclusion}

The widespread adoption of deep neural networks in software
systems has fueled the need for software engineering practices specific to DNNs.
Previous work has shown that like traditional software, DNN software is prone to
bugs albeit with very different characteristics. It is important to further
understand the 
characteristics of bug fixes to inform strategies for repairing 
DNN software that has these bugs. How do developers go about fixing these bugs?
What challenges should automated repair tools address?
To that end, we conducted a comprehensive study to understand 
how bugs are fixed in the DNN software.
Our study has led to several findings. 
First of all, we find that bug fix patterns in DNN are significantly different
from traditional bug fix patterns. 
Second, our results show that fixing the incompatibility between 
the data and the DNN alone can be of significant help to 
developers of DNN software, especially if the developers can be
warned about the impact of their bug fix on the robustness of the
DNN model.
Third, our study shows that a prevalent bug fix pattern is version 
upgrade. While version upgrade is well-studied in SE research,  
our bug fix patterns suggest that automated repair tools will 
need to address at least two unique challenges: invasive, backward
incompatible changes and probabilistic behavior change.
Fourth, our study shows that the structure of the DNN itself 
needs to be represented in repair tools because several fix 
patterns rely on identifying incompatibilities in that structure.
For instance, network connection fixes where disconnected layers
are identified and connected, or adding missing layers, etc.
Fifth, we have found that a significant number of bug fixes introduce new bugs
in the code. 
Finally, we have identified three challenges for fixing bugs:
bug localization is very difficult, reuse of the DNN model is hard because of
limited insights into its behavior, and keeping
up with rapid releases is hard. 

This study opens up several avenues for future work. 
First and perhaps most immediately, a number of bug fix patterns
identified by this work can be automated in repair tools. 
Such tools for bug repairs can help the developers 
integrating DNN into their software.
Second, an abstract representation of the DNN along with the code
that uses it can be developed. We saw several bug fix patterns that
rely on analyzing such a representation. 
Third, there is a critical need to improve bug localization 
for DNN by addressing unique challenges that arise, and by 
creating DNN-aware bug localization tools.
Fourth, there is an urgent need
to detect bugs introduced by dimension mismatch and specially changes that have
the potential to introduce vulnerabilities in
the DNNs. 
Fifth, urgent work is needed on upgrade tools that encode the semantics of
version changes and keep up with the change in the
signature and semantics of DNN libraries. This is important to keep pace with
rapid development in this area. 


\section*{Acknowledgments}

This work was supported in part by US NSF under grants CNS-15-13263, 
and CNS-19-34884. All opinions are of the authors and do not reflect the view of sponsors.
We thank ICSE'20 reviewers for constructive comments that were very helpful.


\balance
\bibliographystyle{ACM-Reference-Format}
\bibliography{refs}


\end{document}